\documentclass[12pt]{article}
\def\beq{\begin{equation}}
\def\eeq{\end{equation}}
\def\beqa{\begin{eqnarray}}
\def\eeqa{\end{eqnarray}}
\def\ban{\begin{eqnarray*}}
\def\ean{\end{eqnarray*}}
\def\bi{\begin{itemize}}
\def\ei{\end{itemize}}

\begin{document}

\begin{center}
{\bf The q-Deformed NJL Model ''Revisited ''}
\end{center}

\begin{center}
{\it S.S. Avancini, J.R. Marinelli, D.P.Menezes and M.M. Watanabe de Moraes}\\
{\it Depto de F\'{\i}sica - CFM -
  Universidade Federal de Santa Catarina}\\
{\it Florian\'opolis - SC - CP. 476 - CEP 88.040 - 900 - Brazil}\\
\end{center}

\vspace{0.50cm}

\begin{abstract}
{In this work we investigate the chiral symmetry breaking in the
$q$--deformed version of the NJL model and its consequent mass
generation mechanism. We show that the deformation of the NJL model, in
the mean field approximation, may take into account correlations that go
beyond the mean field and, in a certain limit, approaches the more
realistic lattice calculations.} 
\end{abstract}

\vspace{0.50cm}
PACS number(s):11.30.Rd: 03.65.Fd: 12.40.-y 
\vspace{0.50cm}

Quantum algebras, or simply {\it q-deformed} algebras, known in the case
where just a single deformation parameter is introduced, have received some
attention last years in the study of many-body problems. Because they can
provide us with a class of symmetries that is richer than the usual Lie
algebras, they are appropriate to describe physical systems and (or) models
which are not properly described by the last ones. On the other hand, the
introduction of the $q$--parameter in the theory can be viewed as a way to
take into account correlations in many-body systems. As examples, we mention
a set of works \cite{JPA94}, \cite{PRA95} related with the deformation of
the su(2) algebra (also called su$_q$(2) algebra) in a time dependent
Hartree--Fock (TDHF) approximation to solve the Lipkin model. Following
similar reasonings, the \textit{quon} algebra, which describes particles
whose statistics interpolates between the bosonic and fermionic ones \cite
{quon1}, was then used to perform a $q$--deformed boson analysis of the
random--phase--approximation (RPA) solution for the Lipkin and the two-level
pairing models \cite{qRPA}. In that case there is a strong indication that
the introduction of correlations through the deformation of the RPA bosons
can also be interpreted as a restoration of a broken symmetry caused by
using the regular (approximated) RPA solution. This is manifested in the
smearing out of the phase transition at the so-called critical point \cite
{RingSchuck}.

Recently, the dynamical breaking of chiral symmetry in the
Nambu-Jona-Lasinio (NJL) model \cite{NJL} and the corresponding generation
of a dynamical mass for quarks, were analyzed at the light of a $q$-deformed
version of that model \cite{celso}. Once $q$-deformation appears as a
powerful tool to study symmetries and their generalizations, this seems to
be a quite appealing investigation. For that purpose, a particular
prescription for the deformation was chosen, based on reference \cite
{Ubriaco} and two important conclusions of that analysis were drawn: the
phase transition of the NJL model continues to be sharp and occur at the
same critical interaction point, when compared with the non--deformed case,
which also means that any possible explicit chiral symmetry breaking terms
in the Lagrangian cannot be simulated by the deformation procedure used; the
main modification caused by the deformation is an enhancement of the
condensate, i.e., the quark mass is effectively increased for a given value
of the interaction constant.

In the present work, it is our intention to proceed further with the
analysis introduced in reference \cite{celso}. We propose a method to solve
the NJL model based on a Hamiltonian density written in terms of the su(2)
operators. As we show bellow, this is in fact possible for the NJL model and
corresponds to a J=1 angular momentum algebra \cite{Japanese}, once we are
restricted to the well-known BCS type ansatz for the variational solution in
the Hartree approximation. With that result in hand, it is then
straightforward to build an su$_q$(2) version of the NJL model. This has the
advantage that it is then possible to exploit different $q$-deformation
schemes \cite{BonnaRev}. Moreover, this procedure allows us to generalize
some previously obtained results and to explore all features introduced by $%
q $-deforming the model.

We start with the Lagrangian density of the NJL model \cite{NJL} 
\begin{equation}
L=\overline{\psi \ }i\partial _\mu \gamma ^\mu \psi +G
\left( (\overline{ \psi }%
\psi )^2+(\overline{\psi }i\gamma _5\tau \psi )^2 \right),  \label{lag}
\end{equation}
where $\psi$ represents the quark field amplitude, the second term
represents the interaction mechanism and $G$ is the coupling constant. From (%
\ref{lag}), the corresponding Hamiltonian density can be obtained and reads: 
\begin{equation}
\mathcal{H}=-i\overline{\psi \ }\mathbf{\gamma .\nabla }\psi -
G \left(( \overline{
\psi }\psi )^2-(\overline{\psi }i\gamma _5\tau \psi )^2 \right). \label{ham1}
\end{equation}

As it is well known, a solution for the above problem can be found in the
Hartree approximation, using a BSC--like variational state \cite{klewansky}
of the form: 
\begin{equation}
|NJL>=\prod_{\mathbf{p}s}\left( \cos \theta (\mathbf{p})-s\sin \theta (%
\mathbf{p})b^{\dagger }(\mathbf{p},s)d^{\dagger }(-\mathbf{p},s)\right)|0>,
\label{varstate}
\end{equation}
where  ${\mathbf p}$ is the vector momentum, s =$\pm 1$ are the helicity state
eigenvalues and the particle and anti--particle operators are defined such
that 
$$b(\mathbf{p},s)|0>=0~~~,~~~d(\mathbf{p},s)|0>=0. $$

In terms of the creation and annihilation operators, we expand the field
operators at t=0, yielding 
$$
\psi (\mathbf{x},0)=\sum\limits_s\int \frac{d^3p}{(2\pi )^3}
\left( b(\mathbf{p},s) u(\mathbf{p},s) e^{i\mathbf{p.x}}+
d^{\dagger }(\mathbf{p},s) v(\mathbf{p},s) e^{-i\mathbf{p.x}} \right), 
$$
where $u(\mathbf{p},s)$ and $v(\mathbf{p},s)$ are the normalized spinor
eigenfunctions for particles and anti--particles with momentum ${\mathbf p}$
and helicity $s$.

At this point,we introduce the following angular momentum operators:
\begin{equation}
J_{+,{\mathbf p}}=\sqrt{2} A_{\mathbf p}^{\dagger }~~,~~J_{-,{\mathbf p}}=
\sqrt{2}A_{\mathbf p}~~,~~
J_{0,{\mathbf p}}=\frac{N_{\mathbf p}}2-1,
\end{equation}
where
\begin{equation}
A_{\mathbf p}^{\dagger }=\frac 1{\sqrt{2}}[b^{\dagger }({\mathbf p},+)
d^{\dagger}(-{\mathbf p},+)-b^{\dagger }({\mathbf p},-)d^{\dagger }
(-{\mathbf p},-)],
\end{equation}
\begin{equation}
N_{\mathbf p}=\sum_s[b^{\dagger }({\mathbf p},s)b({\mathbf p},s)+d^{\dagger }
({\mathbf p},s)d({\mathbf p},s)],
\end{equation}
the operators $J_{+,{\mathbf p}},J_{-,{\mathbf p}}$ and 
$J_{0,{\mathbf p}}$ obey the usual su(2) commutation relations and
\begin{equation}
(J_{+,{\mathbf p}})^n|0>=0,~~ {\rm if}~~ n>2,~~ J_{-,{\mathbf p}}|0>=0,~~
J_{0,{\mathbf p}}|0>=-|0>.
\end{equation}
The variational ansatz defined in equation (\ref{varstate}) can then be 
written as:
\begin{equation}
|NJL>={\mathcal N} e^{-\sum\limits_p\xi _p A_{\mathbf p}^{\dagger }}|0>,
\label{varstate2}
\end{equation}
with
\begin{equation}
{\mathcal N}=\frac 1{\prod\limits_p(1+\frac{|\xi _p|^2}2)},
\end{equation}
where $\xi _p=\sqrt{2} \tan \theta (p)$. In terms of those
operators, the hamiltonian (\ref{ham1}) (in the Hartree approximation) 
can be put in the form:
\begin{equation}
H=\int d^3x\mathcal{H}=\int \frac{d^3p}{(2\pi )^3}\left\{
p~2~J_{0,{\mathbf p}}+M~(J_{+,{\mathbf p}}+J_{-,{\mathbf p}})\right\},
\label{ham2} 
\end{equation}
where $M$ represents the mass of the arising condensate.

We now minimize the mean value of the Hamiltonian with the above NJL state,
using $\theta (p)$ as the variational function and end up with the well
known NJL solution for the condensate, i.e., $<\overline{\psi }\psi >=-M/(2G)
$. At this point we introduce the deformation in the model. Once the
algebraic su(2) structure above underlies the model, there exists a direct
procedure to extend the solution to the su$_q$(2) formalism \cite{BonnaRev}.
The $q$-deformed angular momentum operators obey the following commutation
relations: 
\begin{equation}
[J_{0,{\mathbf p}},J_{\pm ,{\mathbf p}}]=\pm J_{\pm ,{\mathbf p}}~~,~~
[J_{+,{\mathbf p}},J_{-,{\mathbf p}}]=(qr)^{J-J_{0,{\mathbf p}}}  
[2J_{0,{\mathbf p}}],
\end{equation}
where 
$J=\sqrt{ J_{+,{\mathbf p}}^2 +J_{-,{\mathbf p}}^2+J_{0,{\mathbf p}}^2}$
and $[X]$ represents the
deformed version of the operator (or c-number) $X$. Its definition is not
unique. Throughout this paper we use the following definition \cite{Kibler}: 
\begin{equation}
\lbrack X]=\frac{q^X-r^{-X}}{q-r^{-1}},  \label{caixote}
\end{equation}
and only cases where just one deformation parameter is introduced is
considered, namely, $r=q$ and $r=1$, respectively. Note that in the first
case ($r=q$), $q$ can be either a real or a complex number ($q=e^{i\tau
},\tau $ being real), and in the second one ($r=1$) it must be real. To
proceed with the deformed solution we introduce a $q$-deformed variational
ansatz, which is: 
\begin{equation}
|NJL>_q={\mathcal N}_q \emph{e}_q^{-\sum\limits_p \frac{\xi _p}
{\sqrt{2}}J_{+,{\mathbf p}}}|0>,
\label{qvarstate})
\end{equation}
where 
\begin{equation}
{\mathcal N}_q=\frac 1{\prod\limits_p\left( 1+\frac{[2]|\xi _p|^2}2+(\frac
{|\xi _p|^2}2)^2\right) }
\end{equation}
and the $q$--exponential is defined as $e_q(ax)=\sum\limits_{n=0}^\infty 
\frac{a^n}{[n]!}x^n$ , with $[n]!=[n][n-1].....[2][1]$ .

Again, we minimize the mean value of the Hamiltonian (\ref{ham2}), using the
variational state defined in (\ref{qvarstate}) and bearing in mind that the
angular momentum operators obey the deformed commutation rules. After a
straightforward calculation we obtain the modified gap equation: 
\begin{equation}
M=4G[2]\int \frac{d^3p}{(2\pi )^3}\left\{ \frac{\tan \theta (p)(1+\tan
^2\theta (p))}{[1+[2]\tan ^2\theta (p)+\tan ^4\theta (p)]}\right\} ,
\end{equation}
where $\theta (p)$ is a solution of the equation 
\[
\tan ^6\theta (p)+\frac{2p}M\tan ^5\theta (p)+(3-[2])\tan ^4\theta (p)+\frac{%
8p}{[2]M}\tan ^3\theta (p)
\]
\[
-(3-[2])\tan ^2\theta (p)+\frac{2p}M\tan \theta (p)-1=0.
\]

At this point it is useful to make some comments about the above equations.
Our choice for a $q$-deformed exponential in the NJL ansatz is somewhat
arbitrary, in the sense that we could have introduced the deformation just
by modifying the su(2) commutation rules. If we had chosen to write the
variational state (\ref{qvarstate}) in terms of a regular exponential
function, our results would reduce to:

\begin{equation}
M^{*}=4G^{*}\int \frac{{d^3p}}{(2\pi )^3}\left\{ \frac{M^{*}}{\sqrt{%
p^2+(M^{*})^2}}\right\} ,
\end{equation}
where $M^{*}=\sqrt{\frac{[2]}2}M$ and $G^{*}=G[2]$. As $[2]\rightarrow 2$
when $q\rightarrow 1$, this last result also makes clear that we reproduce
the right limit for the gap equation \cite{klewansky}. We have verified
numerically that the difference between the results obtained from the gap
equations (15) and (16) is negligeable, unless we investigate the behavior
of the dynamical mass for $q$ values very far from 1, which is not the case in
the present study. Thus, the introduction of the deformed
exponential, instead of the regular one, in the variational ansatz,
does not affect the results discussed next.

In figures 1a and 1b we show the condensate as a function of the 
inverse coupling constant for the two kinds of deformation considered here, 
i.e., $r=q$ and $r=1$ respectively. In the first case, $q$ can acquire any 
complex values, but
in the second one it must be real. We have chosen two values
in each case in order to show the main modifications introduced by the
deformation. In figure 1a, for $r=q$, we see that for $q$ real we obtain 
an increase
of the condensate, in agreement to what was found in \cite{celso}, and for a
complex $q$ the condensate  decreases, in relation to the condensate obtained
for $q=1$. This is not surprising, once
according to \cite{moszk} and \cite{shelly}, complex $q$-values mimic a
repulsive interaction while real ones play the role of an attractive
interaction. A similar effect can be seen when $r=1$, for which only
real $q$-values are allowed. The results are displayed in figure 1b.
In that case, however, for $q < 1$ a repulsive
effect is simulated, while for $q > 1$ an additional attraction appears.
In all cases we notice that the critical value for the
coupling constant depends on the deformation parameter. 
It can be shown that the phase-transition point is given by 
$G_c^{q}=\frac{2G_c}{[2]}$, with   $G_c=\frac{\pi ^2}{\Lambda ^2}$ where 
$\Lambda $ is the well-known NJL cutoff. Throughout this paper we have used 
$\Lambda =600MeV$. This is a
feature not present in the result obtained in \cite{celso} and we return
to this point latter.

It is also worth mentioning that there is a certain degree of ambiguity in
the way a physical system can be $q$--deformed. This problem has already
been extensively discussed in the literature \cite{nossos,Floratos}. In this
way, equation (\ref{ham2}) can be substituted by 
\begin{equation}
H=\int d^3x\mathcal{H}=\int \frac{ d^3p}{(2\pi )^3}\left\{
p~[2~J_{0,{\mathbf p}}]+M~(J_{+,{\mathbf p}}+J_{-,{\mathbf p}})\right\}.
\label{ham3} 
\end{equation}
before the minimization procedure is performed within the deformation
scheme. As the only modification refers to the way we write the kinetic
term, i.e., we replace $2J_0$ by $[2J_0]$, the corresponding results are called
the deformed kinetic (DK) results and the previous results obtained by
minimization of the Hamiltonian (\ref{ham2}) are called non--deformed kinetic
results (NDK). A comparison between the two approaches is shown in figures 
2a and
2b, for $r=q$ and $r=1$ respectively. In figure 2a, we can see
that the DK results simple turn off de $q$ dependence in G$_c$, yielding the
same behavior obtained in reference \cite{celso}. However, for $r=1$ the DK
results can lead to strong quantitative modifications, specially in what
concerns the phase transition point, as is made clear for $q=2$ in figure 2b.

In order to get some insight on the physical background embedded in the
above results, we have decided to make a qualitative comparison with more 
elaborate
calculations in the NJL model \cite{Khan}. In that work, a
lattice Monte Carlo simulation and a Schwinger-Dyson (S-D) calculation 
are applied to obtain the condensate and compared with the usual gap equation
solution. The first important result that emerges is the fact that 
the value of the critical coupling constant ( G$_c$) depends on the
approximation used. The better the approximation, the bigger the critical
coupling constant. Moreover, both the Monte Carlo simulation and the S-D
method gives a smaller value for the condensate for any given value of the
coupling constant, compared with the gap equation result, as can be seen in
figure 3 of \cite{Khan}. These two features can be simultaneously
reproduced by our $q$-deformed NJL calculation for a complex $q$-value ($r=q$)
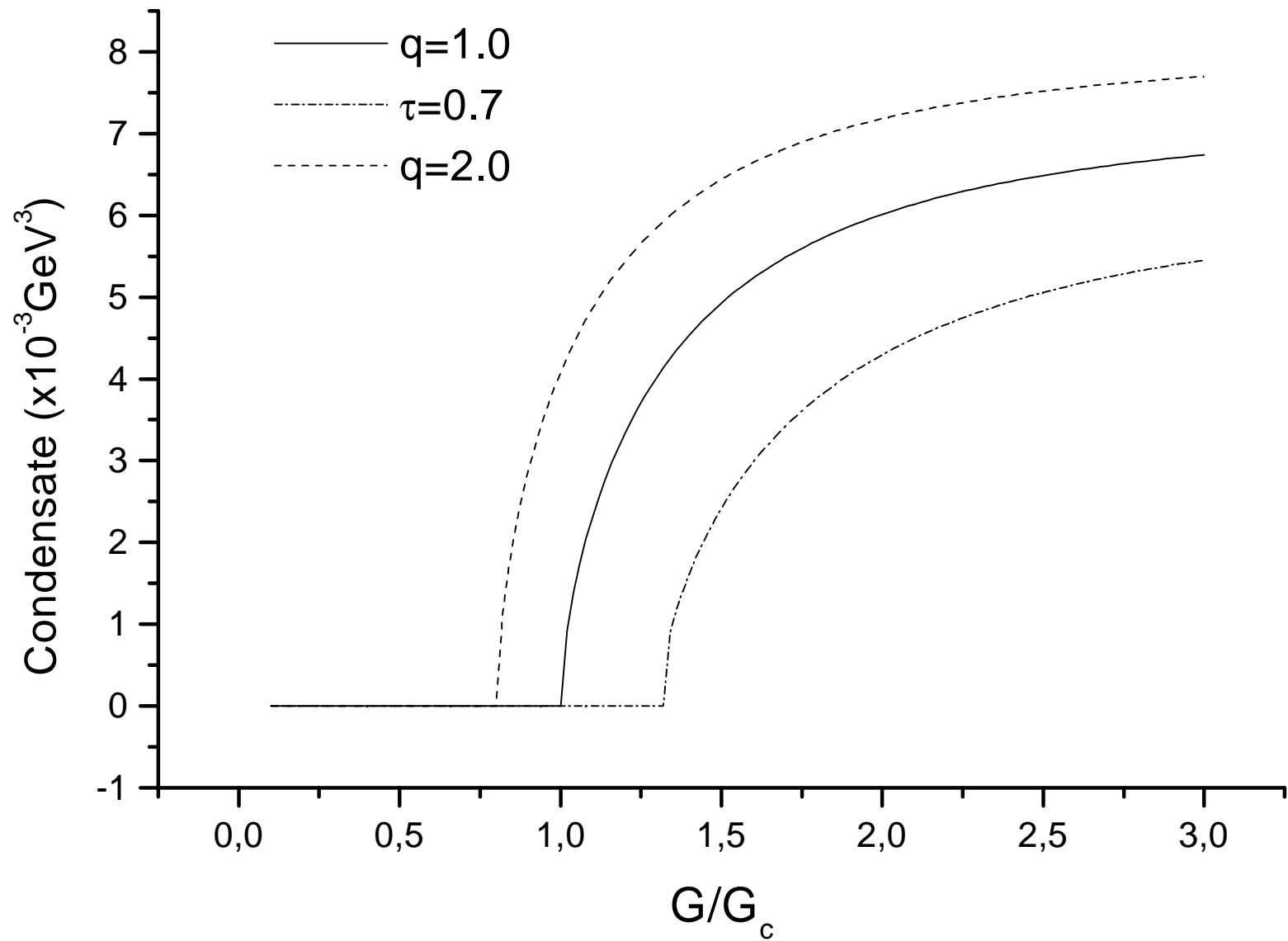
or for $q < 1$ ($r=1$). In figures 3a and 3b we plot the condensate as a
function of $\frac G{G_c}$ for a better comparison with the results in 
\cite{Khan}.

In summary, we have introduced deformation in the NJL model by means of
the angular momentum operators, which are the generators of the $su_q(2)$
algebra. The condensate was then obtained by minimizing the deformed
Hamiltonian with the help of a BCS ansatz for the $q$ variational state. The
results show that the phase transition is never suppressed, but the behavior
of the condensate depends on the definition of the deformed quantity 
and on the numerical value of $q$. Two different
prescriptions to deform the Hamiltonian and two different ways of
introducing the $q$-deformed variational state were used.
 A qualitative comparison was made
with calculations that go beyond the familiar Hartree solution \cite{Khan}.
We can obtain the same behavior as these sophisticated calculations for any
$q$- value which simulates a decrease in the residual attractive interaction. 
Our calculations give rise to a $q$ dependence in the critical strength G$_c.$
This dependence is consistent with the repulsive character that can be mimiced
by the deformation and goes to the right direction as compared with
realistic calculations for the condensate.

\vskip 0.35in

\begin{center}
\textbf{Acknowledgments}
\end{center}

This work was partially supported by CNPq - Brazil.

\end{document}